\documentclass{article}
\usepackage{amssymb,graphicx}

\begin{document}

\newcommand{\beq}{\begin{equation}}
\newcommand{\eeq}{\end{equation}}
\newcommand{\bea}{\begin{eqnarray}}
\newcommand{\eea}{\end{eqnarray}}
\newcommand{\Z}{{\mathbb Z}}
\newcommand{\N}{{\mathbb N}}
\newcommand{\C}{{\mathbb C}}
\newcommand{\Cs}{{\mathbb C}^{*}}
\newcommand{\R}{{\mathbb R}}
\newcommand{\intT}{\int_{[-\pi,\pi]^2}dt_1dt_2}
\newcommand{\cC}{{\mathcal C}}
\newcommand{\cI}{{\mathcal I}}
\newcommand{\cN}{{\mathcal N}}
\newcommand{\cE}{{\mathcal E}}
\newcommand{\cA}{{\mathcal A}}
\newcommand{\xdT}{\dot{{\bf x}}^T}
\newcommand{\bDe}{{\bf \Delta}}

\title{ On the Euclidean Version of the  Photon Number Integral }

\author{ S. Ruijsenaars\\ 
 Department of Applied Mathematics \\
        University of Leeds,\      Leeds LS2 9JT, \ UK \\
\\
L. Stodolsky\\
Max-Planck-Institut f\"ur Physik 
(Werner-Heisenberg-Institut)\\
F\"ohringer Ring 6, 80805 M\"unchen, Germany}

\maketitle

\begin{abstract}
We reconsider the Euclidean version of the photon number integral introduced in \cite{me}. This integral is well defined for any smooth non-self-intersecting curve in $\R^N$.  Besides studying general features of this integral (including its conformal invariance), we
evaluate it explicitly for the ellipse. The result
is $n_{ellipse}=(\xi^{-1}+\xi)\pi^2$, where $\xi$ is the ratio of the
minor and major axes. This is in agreement with the previous result
$n_{circle}=2\pi^2$ and also with the conjecture
that the minimum value of $n$ for any plane curve occurs for the
circle.  
\end{abstract}
  
\section{Introduction}
 The photon number integral \cite{me} encodes the average number of photons radiated by a charged particle moving on a prescribed timelike trajectory in Minkowski space-time.  Its  Euclidean analog is well defined for any smooth closed curve $\cC$ without self-intersections. For such curves we shall study it in terms of a given parametrization
 \beq\label{para}
 {\bf x}\, :\, [-\pi,\pi) \to \R^N,\ \ \ t \mapsto (x_1(t),\ldots,x_N(t)),
 \eeq
 where the coordinates $x_j(t)$ are smooth $2\pi$-periodic functions on the real line, and the tangent $\dot{\bf x}(t)$ does not vanish for any  $t\in [-\pi,\pi)$. 
 The integral is now given by
 \beq\label{nC}
 n_{\cC}=-2 \intT \frac{\xdT (t_1)\cdot \xdT(t_2)}
 {({\bf x}(t_1)-{\bf x}(t_2))^2} .
 \eeq
 Here the ``transverse tangent" $\xdT(t_j)$ stands for the tangent
   at the point ${\bf x}(t_j)$ with the component
along the direction to the other point  removed. (This prescription arises from the transverse polarization of photons~\cite{me}.) That is, if 
\beq
\bDe={\bf x}(t_1)-{\bf x}(t_2)
\eeq
is the difference vector connecting the two points, then
\beq\label{xdT}
\xdT(t_j)=\dot{{\bf x}}(t_j)-(\dot{{\bf x}}(t_j)\cdot \bDe)\bDe/S^2,\ \ \ \ j=1,2,
\eeq
where
\beq\label{defS}
S=|\bDe|.
\eeq
It readily follows from this that $|\xdT(t_j)|$ is proportional to $S(t_1,t_2)$ as $t_1-t_2\to 0$. Hence the integrand remains bounded for  $t_1-t_2\to 0$. It is also easy to see that the integral does not depend on the particular parametrization chosen. Specifically, if
\beq
\phi\, :\, [-\pi,\pi)\simeq S^1 \to [-\pi,\pi),\ \ \  t\mapsto t'
\eeq
is a diffeomorphism of the circle $S^1$, then the integral takes the same value when ${\bf x}(t)$ is replaced by ${\bf x}(\phi(t))$. To express this invariance, one can also write
\begin{equation} \label{two} 
n_{\cC}=- 2\int \int{{\bf dx}^T\cdot {\bf
dx'}^T\over ({\bf x-x'})^2},
\end{equation}
but we shall use (\ref{para}) throughout. 

Some more invariance properties are easily verified, namely, scale invariance and invariance under the Euclidean group (translations and rotations). It is not at all obvious, however, whether the integral is invariant under the inversion
\beq\label{defcI}
\cI \, :\, \R^N\setminus \{ {\bf 0}\}  \to   \R^N\setminus \{ {\bf 0}\},\ \ \  {\bf x}\mapsto {\bf x}/|{\bf x}|^2,
\eeq
assuming the origin is not on the curve. Even so, this is true, as first argued in \cite{inv}. One purpose of this paper is to reconsider this property, for which we supply a rigorous proof.

At this point we would like to mention the two references~\cite{free,druk}, where in particular the conformal invariance of similar integrals is studied. The integrands in these papers differ from that of $n_{\cC}$ (and from each other) in that the divergent behavior of $\dot{\bf x}_1 \cdot \dot{\bf x}_2 /S^2$ for $t_1-t_2\to 0$ is regularized by making different subtractions. 

The main new result of this paper is an explicit evaluation of the integral (\ref{nC}) for the ellipse. Indeed, thus far it was only
possible to evaluate it explicitly for the circle, where it yields the value $2\pi^2$. Some numerical evaluations for the ellipse were given in~\cite{pho}, and these agree with the closed formula,
 \beq\label{nell}
 n_{ellipse}= (\xi^{-1}+\xi)\pi^2,\ \ \ \xi =b/a,
 \eeq
 where $a$ and $b$ are the major and minor axes.

 We present the proof of this formula in Section~2. In Section~3 we reobtain invariance under inversion (hence invariance under the conformal group). In Section~4  we study the behavior of the integral in $\R^N$, $N>2$, when the distance between two parts of the curve goes to 0. Depending on the angle under which a self-intersection develops, we find that the local contribution to the integral can diverge to plus or minus $\infty$. It follows in particular from this result that for $N>2$ there is no lower bound on the integral $n_{\cC}$. Thus $n_{\cC}$ cannot be used to study knots and links in the same way as in~\cite{free}. (The key difference is that the regularization of $\dot{\bf x}_1 \cdot \dot{\bf x}_2 /S^2$ in~\cite{free} yields a positive integrand that diverges to $\infty$ whenever a self-intersection develops.)

\section{The ellipse}

We study the ellipse in the form
\begin{equation}\label{ell}
\frac{x^2}{a^2}+\frac{y^2}{b^2}=1.
\end{equation}
An obvious parametrization of the form (\ref{para}) is then in terms of polar coordinates,
\beq\label{polp}
(x(t),y(t))=(a\cos t,b\sin t).
\eeq

To evaluate  the ellipse integral, it is convenient to employ an alternative way of writing (\ref{nC}) for a curve in $\R^3$. Clearly, this can then be used for any plane curve by embedding $\R^2$ into $\R^3$ in the obvious way, namely by setting $z(t)=0$. The point is that we can apply  the vector identity 
\beq
\bf (a\times
b)\cdot (c\times d)=(a\cdot c)(b\cdot d)-(a\cdot d)(b\cdot c)
\eeq
to obtain
\beq\label{cross}
(\dot{\bf x}_1\times
\bDe)\cdot (\dot{\bf x}_2\times \bDe)=(\dot{\bf x}_1\cdot \dot{\bf x}_2)(\bDe \cdot \bDe )-(\dot{\bf x}_1\cdot \bDe )(\dot{\bf x}_2\cdot \bDe),\ \ \ {\bf x}_j={\bf x}(t_j).
\eeq
Comparing the right-hand-side to (\ref{xdT}), we see it equals $(\xdT_1\cdot\xdT_2)|{\bf \Delta}|^2 $.
Thus (\ref{nC}) can also be written
\begin{equation} \label{tw}
n=- 2\intT  \frac{ {( \dot{\bf x}_1\times\bDe})\cdot (
\dot{\bf x}_2\times\bDe)}{ S^4},\ \ \ \ \ (N=3).
\end{equation}

 Returning to the above ellipse, we have
\begin{equation}\label{v}
\dot{\bf x}_j=(-a\sin t_j,b\cos t_j),\ \ \ \ j=1,2,
\end{equation}
 while for $\bf \Delta$ we get
\bea\label{del}
{\bf \Delta}  &  =  &  \biggl(a(\cos t_1-\cos t_2)
,b(\sin t_1-\sin t_2) \biggr)
\nonumber \\
&  =  &  2\sin({t_1-t_2\over 2})\biggl(-a\,
\sin({t_1+t_2\over
2}),b\,\cos({t_1+t_2\over 2})\biggr),
\eea
where we used well-known trigonometric identities. Thus we obtain
\begin{equation}\label{del2}
S^2=4\sin^2({t_1-t_2\over 2})\biggl( a^2\sin^2({t_1+t_2\over
2})+b^2\cos^2({t_1+t_2\over 2})\biggr),
\end{equation}
with a  factorization in terms of sum and difference angles.

Since all vectors are in the $(x,y)$-plane, the cross products are 
in the $z$-direction. In particular,  from (\ref{polp}) we deduce
\begin{equation} \label{am} 
{\bf x}\times \dot{\bf x}=ab {\bf \hat z}.
\end{equation}
In our parametrization, therefore, the ``angular momentum'' (with
respect to the origin---not to the focus of the ellipse) is
constant. Likewise, we obtain
\begin{equation} \label{1cross} 
 \dot{\bf x}_1\times \bDe=-ab\bigl(1-\cos(t_1-t_2)\bigr){\bf \hat
z},\ \ \ \ \ \ 
\dot{\bf x}_2\times \bDe=ab\bigl(1-\cos(t_1-t_2)\bigr){\bf \hat z}.
\end{equation}
Therefore the lhs of (\ref{cross}) becomes
\begin{equation} \label{crossb} 
(\dot{\bf x}_1\times
\bDe)\cdot (\dot{\bf x}_2\times \bDe)=-a^2b^2 (1-\cos(t_1-t_2))^2=-4a^2b^2 \sin^4({t_1-t_2\over
2}).
\end{equation}
Substituting this and (\ref{del2}) in  (\ref{tw}), we finally obtain
\begin{equation} \label{twob} 
n= {a^2 b^2\over 2}\int_{[-\pi,\pi]^2}
 {dt_1dt_2\over \bigl( a^2\sin^2({t_1+t_2\over 2})+b^2\cos^2({t_1+t_2\over
2})\;\bigr)^2 } .
\end{equation}

It remains to calculate the integral on the rhs of (\ref{twob}). Its integrand is smooth and $2\pi$-periodic in $t_1$ and $t_2$, so we can transform to sum and difference variables to get
\beq\label{nJ}
n=a^2b^2\pi J,
\eeq
where
\beq
J=\int_{-\pi}^{\pi}\frac{dx}{(a^2\sin^2x+b^2\cos^2x)^2}.
\eeq
The integral $J$ can be readily calculated via integrals we have occasion to invoke in Section~4, too.  We first note it can be rewritten as
\beq\label{JT}
J=\int_0^{2\pi}
\frac{dx}{(\beta -\alpha \cos(2x))^2}=-\partial_{\beta}T(\beta),
\eeq
where
\beq\label{ab}
\alpha =(a^2-b^2)/2,\ \ \ \beta=(a^2+b^2)/2,
\eeq
\beq\label{defTb}
T(\beta)=\int_0^{2\pi}
\frac{du}{(\beta -\alpha \cos u)}, \ \ \ \beta>\alpha>0.
\eeq
Now $T(\beta)$ can be calculated by a contour integration, the result being
\beq\label{Tev}
T(\beta)=2\pi (\beta^2-\alpha^2)^{-1/2}.
\eeq
From (\ref{JT}) and (\ref{ab}) we then infer
\beq
J=\pi (a^2+b^2)/a^3b^3,
\eeq
and substituting this in (\ref{nJ}) we obtain the explicit result
\beq\label{nf}
n=(\xi +\xi^{-1})\pi^2,\ \ \ \xi=b/a.
\eeq

This is the new result (\ref{nell}) announced above. We proceed to comment on its features. It 
 depends  on the
dimensionless ratio $\xi$, and also exhibits   symmetry under 
$\xi\to \xi^{-1}$, i.e., the exchange of $a$ and $b$. The dependence on
the ratio
reflects the dimensionless character of $n$ and that it only depends
on the
shape of a curve and not on its absolute scale.
 The exchange of $a$ and $b$ is 
equivalent to the rotation of the ellipse by $90^{\rm o}$ and so 
invariance under this exchange
 was to be expected  on grounds of rotational invariance.

It is a plausible conjecture that for any plane curve 
 the minimum value of $n$ 
 obtains for the circle, and it is indeed true that the minimum of 
(\ref{nf}) holds for $\xi=1$. The indefinite
increase in $n$ as the 
ellipse becomes more and more eccentric manifests the increased
``radiation'' from
the acute ends. (We stress, however, that since here we are in
Euclidean space the
notion of ``photon'' should not be taken too literally.)

Finally, in \cite{pho} $n$ was evaluated  numerically for 
ellipses of various 
eccentricities~$\epsilon$. Taking into account that
$\xi=(1-\epsilon^2)^{1/2}$, we find
agreement between the numerical calculations and (\ref{nf}).

\section{Inversion revisited}

Consider a circle $C$ in $\R^2$ with center $c=(x_0,y_0)$ and radius $R$, which does not pass through the origin. It is given by the equation
\beq\label{ce}
(x-x_0)^2+(y-y_0)^2=R^2,
\eeq
with the difference
\beq
d=R^2-x_0^2-y_0^2,
\eeq
being non-zero. Substituting 
\beq\label{inv2}
x\to x/(x^2+y^2),\ \ \ y\to y/(x^2+y^2),
\eeq
in (\ref{ce}), we obtain the equation
\beq\label{cei}
d(x^2+y^2)+2x_0x+2y_0y-1=0.
\eeq
In the degenerate case $d=0$ this yields a line. For $d\ne 0$ one easily verifies that (\ref{cei}) describes a circle $\tilde{C}$ with center $\tilde{c}$ and radius $\tilde{R}$ given by
\beq
\tilde{c}=-d^{-1}(x_0,y_0),\ \ \ \tilde{R}=R|d|^{-1}.
\eeq

The upshot of this calculation is that the inversion of the circle $C$ yields a new circle $\tilde{C}$, assuming $C$ does not pass through the origin. In this case, therefore, one obtains the same value $2\pi^2$ for the curve and its image under inversion. 

Next, consider the ellipse $E$ given by (\ref{ell}) with $a>b$.  Its image $\tilde{E}$ under the inversion (\ref{inv2}) is given by the equation
\beq\label{lima}
\frac{x^2}{a^2}+\frac{y^2}{b^2}=(x^2+y^2)^2,\ \ \ \ a>b>0.
\eeq
Thus the new curve $\tilde{E}$ is not an ellipse. Rather, (\ref{lima}) yields a so-called lima\c{c}on. At face value, it would seem untractable to calculate $n$ for this curve, but in fact this can be done, the answer being the same as for the ellipse $E$!

The reason why this is true is that this amounts to a special case of the invariance under inversion for any curve $\cC$ in the class defined in the introduction, it being assumed in addition that $\cC$ does not pass through the origin. This striking property was first observed in~\cite{inv}. We proceed to reobtain this feature rigorously, the analysis in~\cite{inv} being a bit formal. The key algebraic identities can be found in~\cite{inv}, however: In our given parametrization (\ref{para}) the first one reads
\beq\label{id1}
-\partial_{t_1}\partial_ {t_2}\ln S -(\partial_{t_1}\ln S)(\partial_{t_2}\ln S)=( \xdT_1\cdot\xdT_2)/S^2,
\eeq
where $S$ is given by (\ref{defS}). Its validity can be established by a straightforward calculation, using in particular
\beq\label{part}
\partial_{t_1}\ln S^2=2(\bDe\cdot \dot{\bf x}_1)/S^2,\ \ \ \partial_{t_2}\ln S^2=-2(\bDe\cdot \dot{\bf x}_2)/S^2.
\eeq
 The point of the identity (\ref{id1}) is that it expresses the integrand of $n$ as a function of $\ln S$ only. Thus one need only note the behavior of $\ln S$ under the inversion $\cI$ (cf.~(\ref{defcI})),  
\beq
\ln S \to \ln S -\frac{1}{2}\sum_{j=1}^2\ln ({\bf x}_j\cdot {\bf x}_j),
\eeq
and calculate $t_j$-partials to get
\beq
\partial_1\partial_2 \ln S\to  \partial_1\partial_2 \ln S,
\eeq
\beq
\partial_j \ln S \to \partial_j \ln S -({\bf x}_j\cdot \dot{\bf x}_j)/|{\bf x}_j|^2.
\eeq
From the identity (\ref{id1}) one then concludes
\beq
( \xdT_1\cdot\xdT_2)/S^2\to ( \xdT_1\cdot\xdT_2)/S^2 +I,
\eeq
where
\beq
I=-\frac{({\bf x}_1\cdot \dot{\bf x}_1)}{|{\bf x}_1|^2}\frac{({\bf x}_2\cdot \dot{\bf x}_2)}{|{\bf x}_2|^2} +
\frac{({\bf x}_1\cdot \dot{\bf x}_1)}{|{\bf x}_1|^2}\partial_2 \ln S +\frac{({\bf x}_2\cdot \dot{\bf x}_2)}{|{\bf x}_2|^2}\partial_1 \ln S.
\eeq
The upshot is that the invariance of $n$ under inversion is equivalent to vanishing of the $I$-integral~\cite{inv}.

Now the first term in $I$ is of the form $-f'(t_1)f'(t_2)/4$, where
\beq\label{deff}
f(t)=\ln ({\bf x}(t)\cdot {\bf x}(t)).
\eeq
Since $|{\bf x}(t)|$ does not vanish by assumption, 
$f(t)$ is a smooth $2\pi$-periodic function. Hence it is immediate that the integral of the first term vanishes.

For the second and third term one would also be inclined to integrate the partials of $\ln S$ directly. However, the analytical difficulty is that  they both have a singularity for $t_1-t_2\to 0$, so that one cannot integrate these terms without further ado. On the other hand, their sum is a continuous function on $[-\pi,\pi]^2$,  as we shall show shortly. Therefore the integral of the sum can be obtained as the limit of an integral where an $\epsilon$-neighborhood of the ``diagonal" is omitted. In the latter integral we are entitled to integrate directly, and then take $\epsilon$ to 0. From suitable bounds it then follows that the limit vanishes.

We proceed to fill in the details of this reasoning. To this end we need only consider the remaining integrand
\beq
R(t_1,t_2) = f'(t_1)K_2(t_1,t_2)+f'(t_2)K_1(t_1,t_2),\ \ \ \ \ t_1,t_2\in [-\pi,\pi],
\eeq
where  $K$ is given by
\beq\label{defK}
K=\ln S^2,
\eeq
and $K_j$ stands for $\partial_jK$.
To begin with, we have expansions
\beq
\bDe (t,t+\epsilon)=-\epsilon \dot{\bf x}-\epsilon^2\ddot{\bf x}/2+O(\epsilon^3),
\eeq
\beq\label{expS}
S^2(t,t+\epsilon)=\epsilon^2\dot{\bf x}^2+\epsilon^3\dot{\bf x}\cdot \ddot{\bf x}+O(\epsilon^4).
\eeq
Using (\ref{part}) this entails
\bea
K_1  &  =  &  \frac{2\dot{\bf x}\cdot (-\epsilon \dot{\bf x}-\epsilon^2\ddot{\bf x}/2 +O(\epsilon^3))}
{\epsilon^2\dot{\bf x}^2+\epsilon^3\dot{\bf x}\cdot \ddot{\bf x}+O(\epsilon^4)}
\nonumber \\
  &  =  &  -\frac{2}{\epsilon}+\frac{\dot{\bf x}\cdot \ddot{\bf x}}{\dot{\bf x}\cdot \dot{\bf x}} +O(\epsilon),
  \eea
  and likewise
  \beq
K_2 = \frac{2}{\epsilon}+\frac{\dot{\bf x}\cdot \ddot{\bf x}}{\dot{\bf x}\cdot \dot{\bf x}} +O(\epsilon).
\eeq
Thus we obtain
\beq
R(t,t+\epsilon)=-2f''(t)+2f'(t)\frac{\dot{\bf x}\cdot \ddot{\bf x}}{\dot{\bf x}\cdot \dot{\bf x}} +O(\epsilon).
\eeq
Hence $R$ is continuous on all of the integration region $[-\pi,\pi]^2$. (Recall we are dealing with smooth $2\pi$-periodic functions $x_1(t),\ldots,x_N(t)$, so that the above expansions also apply to $\epsilon$-neighborhoods of $t_1=-\pi,t_2=\pi$ and $t_1=\pi,t_2=-\pi$, where $\bDe$ vanishes too.)

To prove that $R$ has vanishing integral, it therefore suffices to show
\beq
\lim_{\epsilon\to 0}R_{\epsilon}=0,
\eeq
where
\beq\label{Reint}
R_{\epsilon}=\int_{J_{\epsilon}}dt_1dt_2R(t_1,t_2),
\eeq
\beq
J_{\epsilon}=[-\pi,\pi]^2\setminus \{ t_2\in (-\pi +\epsilon, \pi -\epsilon), |t_1-t_2|<\epsilon\}.
\eeq
Thus we are not only excising $\epsilon$-neighborhoods of the diagonal and the corners $(-\pi,\pi)$ and $(\pi,-\pi)$ of the $(t_1,t_2)$-square, but also two horizontal strips of height $\epsilon$, cf.~Fig.~1. These
strips allow us to avoid dealing
with the corners, and to choose the order of
the integrations depending on which of the two terms in
$R(t_1,t_2)$ is in question.

Specifically, we rewrite the integral (\ref{Reint}) as
\bea
R_{\epsilon}  &  =  &  \int_{-\pi+\epsilon}^{\pi-\epsilon}dt_2\int_{-\pi}^{t_2-\epsilon}dt_1f'(t_2)K_1+\int_{-\pi}^{\pi-2\epsilon}dt_1\int_{t_1+\epsilon}^{\pi-\epsilon}dt_2f'(t_1)K_2
\nonumber\\
  &   & +\int_{-\pi+\epsilon}^{\pi -\epsilon}dt_2\int_{t_2+\epsilon}^{\pi}dt_1f'(t_2)K_1+\int_{-\pi+2\epsilon}^{\pi}dt_1\int_{-\pi+\epsilon}^{t_1-\epsilon}dt_2f'(t_1)K_2.
  \eea
  The first line refers to the upper left area of the central square in Fig.~1
and the second line to the lower right area.
We  can now  do the inner integrations to get
\bea
R_{\epsilon}  &  =  &  \int_{-\pi+\epsilon}^{\pi-\epsilon}dt f'(t)[K(t-\epsilon,t)-K(-\pi,t)]
\nonumber\\
&  &  +
\int_{-\pi}^{\pi-2\epsilon}dtf'(t)[K(t,\pi -\epsilon)-K(t,t+\epsilon)]
\nonumber\\
&   & +\int_{-\pi+\epsilon}^{\pi -\epsilon}dtf'(t)[K(\pi,t)-K(t+\epsilon,t)]
\nonumber\\
&  &  +
\int_{-\pi+2\epsilon}^{\pi}dtf'(t)[K(t,t-\epsilon)-K(t,-\pi+\epsilon)].
\eea
Rearranging, we obtain
\bea\label{Rfin}
R_{\epsilon}  &  =  &  \int_{-\pi+\epsilon}^{\pi-\epsilon}dt f'(t)[K(t-\epsilon,t)-K(t+\epsilon,t)]
\nonumber\\
&  &  +
\int_{-\pi +2\epsilon}^{\pi-2\epsilon}dtf'(t)[K(t,\pi -\epsilon)-K(t,-\pi+\epsilon)+K(t,t-\epsilon)-K(t,t+\epsilon)]
\nonumber\\
&  &  +
\int_{-\pi}^{-\pi+2\epsilon}dtf'(t)[K(t,\pi -\epsilon)-K(t,t+\epsilon)]
\nonumber\\
&  &  +
\int_{\pi-2\epsilon}^{\pi}dtf'(t)[K(t,t-\epsilon)-K(t,-\pi+\epsilon)].
\eea
The first integral results from combining the first and third lines of the
previous equation. In the second integral we have arranged for symmetric
limits of integration, while the last two integrals are only over $\epsilon$-regions. 

It remains to estimate the various terms occurring in (\ref{Rfin}). The integration region of the terms on the third and fourth line has measure $2\epsilon$, whereas the integrand is $O(\ln(1/\epsilon))$ on the region, cf.~(\ref{defK})--(\ref{expS}). Thus the $\epsilon\to 0$ limit of these terms vanishes. Also, from (\ref{expS}) and a similar expansion we deduce
\beq
K(t-\epsilon,t)-K(t,t+\epsilon)=\ln\left( \frac{\epsilon^2\dot{\bf x}^2+O(\epsilon^3)}
{\epsilon^2\dot{\bf x}^2+O(\epsilon^3)}\right) =O(\epsilon),
\eeq
\beq
K(t,t-\epsilon)-K(t,t+\epsilon)=O(\epsilon),
\eeq
and from
\beq
{\bf x}(\pi-\epsilon)={\bf x}(\pi)-\epsilon \dot{\bf x}(\pi)+O(\epsilon^2),\ \ 
{\bf x}(-\pi+\epsilon)={\bf x}(\pi)+\epsilon \dot{\bf x}(\pi)+O(\epsilon^2),
\eeq
we get
\beq
K(t,\pi-\epsilon)
-K(t,-\pi+\epsilon)=O(\epsilon).
\eeq
Recalling $f'(t)$ is smooth, it now follows that the limits of the terms on the first and second lines also vanish. Thus the $I$-integral vanishes. We have therefore completed our proof of the invariance of $n_{\cC}$ under inversion, assuming the curve $\cC$ does not pass through the origin (the center of inversion).
\begin{figure}
\includegraphics[width=\hsize]{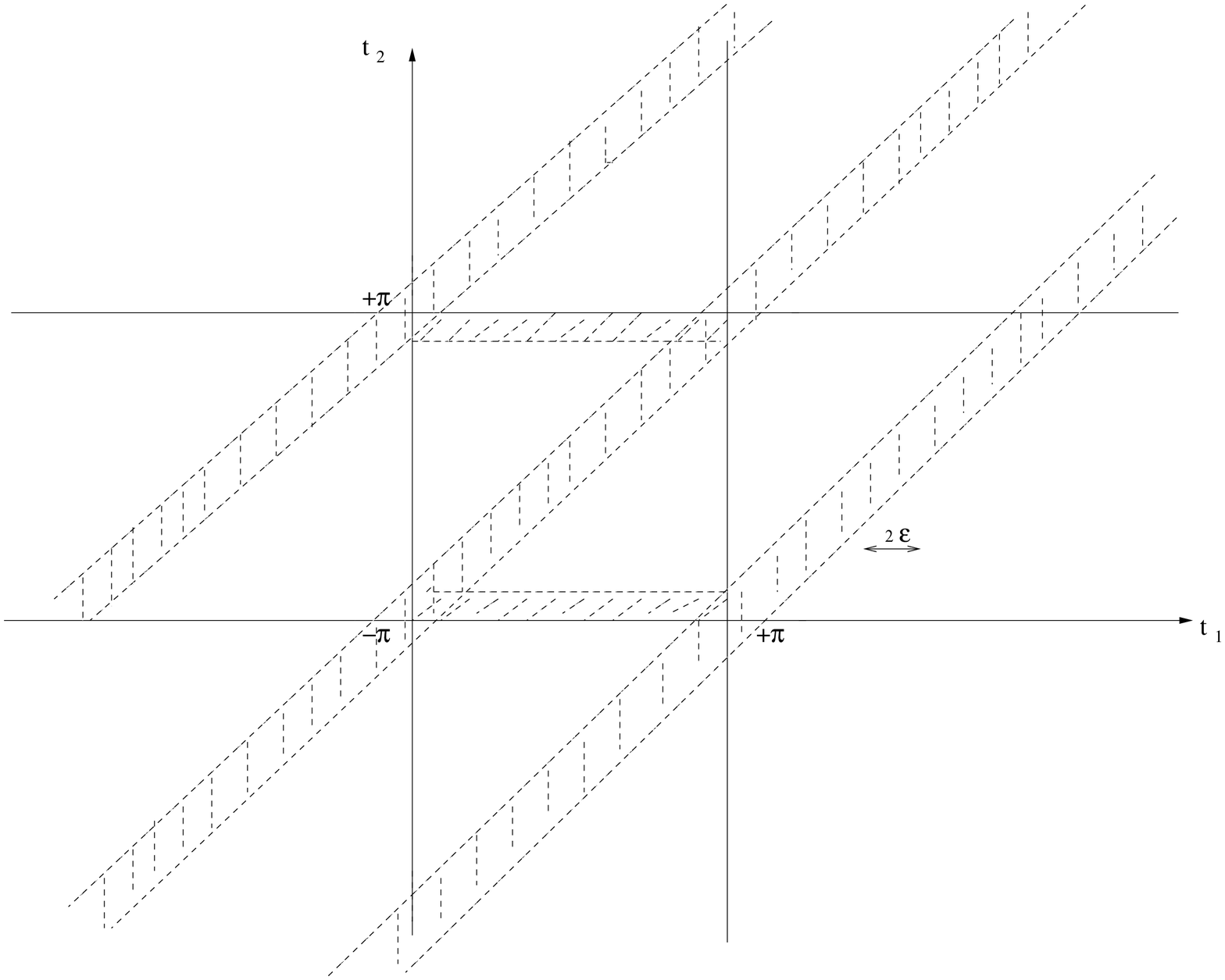}
\caption{ The $(t_1,t_2)$-plane, showing excised areas. These include a
$2\epsilon$-wide region along the $t_1=t_2$ line and its $2\pi$-periodic repetitions,
as well as two $\epsilon$-high strips in the central square. The
integration is only over the central square.}
\end{figure}

%\newpage

\section{The intersection behavior}

In this section we take $N>2$ and study what happens when we let the distance $d$ between two parts of the non-self-intersecting curve $\cC$ go to zero. Since the tangents to the two parts are already transverse at the points of minimal distance, their inner product does not generally vanish as $d$ goes to zero. But the distance $S=d$ does go to zero,  so the integrand of $n$ diverges for the two points of closest approach. 

Of course, this is a priori compatible with the contribution of the two parts to the $n$-integral remaining finite. Here we analyze what happens under the simplifying assumption that the angle $\phi$ between the tangents
at the points $P_1$ and $P_2$ of closest approach 
 remain constant as the distance $d$ between $P_1$ and $P_2$ vanishes. Since we are dealing with smooth curves and we are studying a local behavior, we may also assume that the two parts are straight. Finally, using a change of parameters and an eventual rotation and translation, we may and will study the case of two line pieces in $\R^3$ given by
\beq
{\bf x}_1= (t_1\cos\phi,t_1\sin\phi,d),\ \ \ \ {\bf x}_2=(t_2,0,0),\ \ \ \ t_1,t_2\in [-t_0,t_0],
\eeq
so that $|t_1|$ and $|t_2|$ are the distances from 
\beq
P_1=(0,0,d),\ \ \ \ P_2=(0,0,0).
\eeq

Since we now work in $\R^3$, we can use the vector product form of the integrand featuring in (\ref{tw}). As we have
\beq
\dot{\bf x}_1\times \bDe = (d\sin\phi, d\cos\phi, t_2\sin\phi),\ \ \dot{\bf x}_2\times \bDe = (0,d,t_1\sin\phi),
\eeq
this readily yields 
\beq
\frac{t_1t_2\sin^2\phi+d^2\cos\phi}{(t_1^2+t_2^2-2t_1t_2\cos\phi+d^2)^2},\ \ \ \ t_1,t_2\in [-t_0,t_0].
\eeq

Now we have two distances $d$ and $t_0$ in our problem, and the $n$-integral is dimensionless (scale-invariant). Therefore, the latter can only depend on the ratio
\beq
\mu =d/t_0.
\eeq
Indeed, from the change of variables
\beq
t=t_1/t_0,\ \ \ \ t'=t_2/t_0,
\eeq
we get the integrand
\beq
M_{\mu,\phi}(t,t')=\frac{tt'\sin^2\phi+\mu^2\cos\phi}{(t^2+t'^2-2tt'\cos\phi+\mu^2)^2},\ \ \ \ t,t'\in [-1,1].
\eeq
We are concerned with the divergence or convergence behavior of the integral of $M_{\mu,\phi}$ over $[-1,1]^2$ as $\mu$ vanishes, and since the only possible divergence of the integrand arises from the origin, we may as well study the integral over the unit disk in the $(t,t')$-plane. The crux of this is that we can then pass to polar coordinates and use integrals occurring in Section~2 to do the angular integration explicitly. This yields an integral over the radius whose behavior for $\mu\to 0$ can be easily determined. The details now follow.  

To start with, the integral over the unit disk is given by
\beq\label{unit}
\cI(\mu,\phi)= \int_0^1  r \cA(r)dr,
\eeq
where $\cA$ is the angular integral
\beq\label{cA}
\cA(r)=\int_0^{2\pi}d\theta\frac{r^2 \cos\theta\sin\theta\sin^2\phi+\mu^2\cos\phi}{[r^2(1-2\cos\theta\sin\theta\cos\phi)+\mu^2]^2}.
\eeq
Next we rewrite (\ref{cA}) as
\beq
\cA(r)=\frac{1}{2}\int_0^{2\pi}dx\frac{p\cos x+q}{(\beta-\alpha\cos x)^2},
\eeq
where we have introduced
\beq\label{newp}
p=r^2\sin^2\phi,\ \ q=2\mu^2\cos\phi,\ \ \alpha=r^2\cos\phi,\ \ \beta =r^2+\mu^2.
\eeq
Recalling (\ref{defTb}), we now see that we have
\beq
\cA(r)=-\frac{1}{2\alpha}[pT(\beta)+(p\beta +q\alpha)\partial_{\beta}T(\beta)].
\eeq
It is easy to check that the evaluation (\ref{Tev}) of $T(\beta)$ is also valid for $\alpha\in (-\beta,0]$. Using it, we get
\beq
\cA(r)= \frac{\pi(p\alpha +q\beta)}{(\beta^2-\alpha^2)^{3/2}}.
\eeq
If we now substitute (\ref{newp}) and simplify the result, we obtain
\beq
\cA(r)=\pi\cos\phi [L(r)^{-1/2}+\mu^4L(r)^{-3/2}],
\eeq
where
\beq
L(r)=r^4\sin^2\phi+2\mu^2r^2+\mu^4.
\eeq

The upshot is that the unit disk integral (\ref{unit}) is given by
\beq
\cI(\mu,\phi)=\frac{\pi}{2}\cos\phi \int_0^{1/\mu^2}\left( \frac{du}{(u^2\sin^2\phi+2u+1)^{1/2}}
+\frac{du}{(u^2\sin^2\phi+2u+1)^{3/2}}\right).
\eeq
Thus for $\mu\to 0$ we get the asymptotic behavior
\beq
\cI(\mu,\phi)\sim \pm 2^{-1/2}\pi /\mu,\ \ \ \ \ \cos\phi =\pm 1,
\eeq
\beq
\cI(\mu,\phi)\sim \pi \frac{\cos\phi}{|\sin\phi|}\ln\left(\frac{1}{\mu}\right),\ \ \ \ \phi\ne 0,\pi.
\eeq
As a consequence, we obtain divergence to $\infty$ for $|\phi|\in [0,\pi/2)$ and divergence to $-\infty$ for $|\phi|\in (\pi/2,\pi]$. Note that it is already clear from (\ref{cA}) that when the line pieces are orthogonal the local contribution to $n$ vanishes identically in $\mu$.

\vspace{1.0 cm}

\noindent
{\Large\bf Acknowledgments}

\vspace{1.0 cm}

\noindent
The results reported in this paper were obtained during a stay of S.~R. at the 
Max-Planck-Institute for Physics in Munich (Heisenberg
Institute). He would like to thank the Institute for its hospitality and financial support,
and E.~Seiler for his
invitation and for useful discussions.

%\newpage

\end{document}